\documentclass[aps,prd
,nofootinbib,showpacs
,floatfix,superscriptaddress,preprintnumbers]{revtex4}
\usepackage{amsmath}
\usepackage{amssymb}
\usepackage{epsfig}
\usepackage{graphicx}
\usepackage{stmaryrd}
\usepackage{pstricks}
\usepackage{pst-coil}
\DeclareMathAlphabet{\mathsc}{OT1}{cmr}{m}{sc}

\renewcommand{\baselinestretch}{1.14}

\newcommand{\beq}{\begin{equation}}
\newcommand{\eeq}{\end{equation}}
\newcommand{\bea}{\begin{eqnarray}}
\newcommand{\eea}{\end{eqnarray}}
\newcommand{\bec}{\begin{center}}
\newcommand{\eec}{\end{center}}
\newcommand{\bei}{\begin{itemize}}
\newcommand{\eei}{\end{itemize}}

\newcommand{\nn}  {\nonumber}

\def\10{$SO(10)$}
\def\21{SU(2) $\otimes$ U(1) }

\def\422{$SU(4) \otimes SU(2) \otimes SU(2)$}
\def\321{SU(3) $\otimes$ SU(2) $\otimes$ U(1)}

\def\lsim{\raise0.3ex\hbox{$\;<$\kern-0.75em\raise-1.1ex\hbox{$\sim\;$}}}
\def\gsim{\raise0.3ex\hbox{$\;>$\kern-0.75em\raise-1.1ex\hbox{$\sim\;$}}}
\def\vev#1{\left\langle #1\right\rangle}
\def\eq#1{eq.~(\ref{#1})}

\newcommand{\AddrAHEP}{%
  AHEP Group, Institut de F\'{\i}sica Corpuscular --
  C.S.I.C./Universitat de Val{\`e}ncia \\
  Edificio Institutos de Paterna, Apt 22085, E--46071 Valencia, Spain}

\renewcommand{\baselinestretch}{1.1}
 \newcommand{\ba}{\begin{array}}
\newcommand{\ea}{\end{array}}
\relax

\def\321{$SU(3)\times SU(2)\times U(1)$}

\begin{document}

\preprint{IFIC/08-58}

\title[ $S_4$  as a natural flavor symmetry for lepton mixing]{$S_4$  as a natural flavor symmetry for lepton mixing}
\author{Federica Bazzocchi and   Stefano Morisi \\
\vspace{2mm}
\it{\AddrAHEP}}

\date{\today}

\begin{abstract}
Group theoretical arguments seem to indicate the 
discrete symmetry $S_4$ as the minimal 
flavour symmetry compatible with tribimaximal neutrino mixing. We prove in a model independent way that indeed $S_4$ can realize exact tribimaximal  mixing through different symmetry breaking patterns. We present   two  models in which lepton tribimaximal mixing is realized in different ways and for each one we discuss the superpotential that leads to the correct breaking of the flavor symmetry.
\end{abstract}
\maketitle

\section{Introduction}
Harrison, Perkins and Scott (HPS)  \cite{Harrison:2002er} proposed the so called tribimaximal mixing matrix
\begin{equation}
\label{eq:HPS}
U_{TB} = 
\begin{pmatrix}
\sqrt{2/3} & 1/\sqrt{3} & 0\\
-1/\sqrt{6} & 1/\sqrt{3} & -1/\sqrt{2}\\
-1/\sqrt{6} & 1/\sqrt{3} & 1/\sqrt{2}
\end{pmatrix}.
\end{equation}
This matrix keeps in surprising  agreement with experimental data \cite{Maltoni:2004ei}. 
Lot of theoretical models has been done to explain the mixing matrix of  \eq{eq:HPS} 
by means of non abelian flavor symmetry, such as 
$S_3$\cite{Feruglio:2007hi,Grimus:2005mu,Mitra:2008bn,Mondragon:2007jx,Chen:2007zj,Koide:2006vs,Mohapatra:2006pu,Morisi:2005fy,
Caravaglios:2005gw,Harrison:2003aw,Chen:2004rr}, 
$A_4$ 
\cite{Ma:2001dn,Babu:2002dz,Hirsch:2003dr,Ma:2004zv,Altarelli:2005yp,Chen:2005jm,Zee:2005ut,Altarelli:2005yx,
Adhikary:2006wi,Valle:2006vb,Adhikary:2006jx,Ma:2006vq,Altarelli:2006kg,Hirsch:2007kh,Altarelli:2007cd,Bazzocchi:2007na}, 
$T'$ \cite{Feruglio:2007uu,Aranda:2007dp,Eby:2008uc,Frampton:1994rk,Ding:2008rj},
$S_4$ \cite{Ma:2005pd,Hagedorn:2006ug,Cai:2006mf,Zhang:2006fv,Koide:2007sr} and 
$\Delta(27)$ \cite{Luhn:2007uq,deMedeirosVarzielas:2006fc,Ma:2007wu,Ma:2006ip}.
The non abelian  discrete groups   have irreducible representations of 
dimension bigger than one \cite{Frampton:2000mq}. The most interesting case arises when the group contains  a   triplet as  irreducible representation, allowing to 
embed the observed three generations of fermions.

When  a  non abelian discrete  group $G$   is  broken to one of its subgroup $G'$    the  transformation  $U_{G'}$ that decomposes the representations  of $G$ according to $G'$ can be  fixed and are completely model independent. This is the case for example of $A_4$ broken to $Z_3$: the triplet representation of $A_4$ is sent to the one-dimensional representations of $Z_3$,  $1,1',1''$, through the matrix $U_\omega$ defined as
\bea
\label{uom0}
U_\omega&=&\frac{1}{\sqrt{3}} \left(\begin{array}{ccc} 1&1 &1\\ 1&\omega & \omega^2 \\ 1 & \omega^2 &\omega  \end{array} \right)\,,
\eea
while the one dimensional representations of $A_4$ coincide with the corresponding ones of $Z_3$.  A good candidate to give TBM is  a discrete group $G$ that has  a triplet representation,  at least two subgroups,  $G'$ that decompose according to $U_{G'}$ and $G''$ that decompose according to $U_{G''}$.  
It is necessary having at least  two different subgroups of $G$  to obtain a lepton mixing matrix different to the identity:  if $G$ were  broken to the same subgroup $G'$  both in the charged lepton and in the neutrino sector the lepton mixing matrix would be given by
$ U_{lep}= U^\dag_{G'} \, U_{G'} =\mathcal{I} \,.$

\emph{A priori} $A_4$ seems to be a good candidate because it  is the smallest discrete group  that contains a  triplet as  irreducible representation.  Furthermore it has  two different  subgroups, $Z_3$ and $Z_2$. However, while   the transformation associated to $Z_3$ is  given by  $U_\omega$ the one associated to $Z_2$ is model dependent. 
This  analysis has been already  performed in \cite{Bazzocchi:2008sp} (see eq.~A4).   A similar analysis done  with the discrete symmetry $T'$ lead to the same conclusion (see eq. (8) of Ref.\cite{Feruglio:2007uu}). 
This means that  $A_4$ and $T'$ yield exact  or approximate TBM only assuming  
 a fine tuning in the parameters of the Yukawa lagrangian or a particular model realization.  We mention that by assuming further constraints, also  models based on  $S_3$ can yield  an approximate TBM, although its largest irreducible representation is a doublet and not a triplet.

It has been recently claimed \cite{Lam:2008sh} that the minimal flavor symmetry 
naturally related to the tribimaximal mixing is 
$S_4$, the permutation symmetry of four objects.
 The author of 
\cite{Lam:2008sh} proved this through  group theoretical  arguments without entering into the details of a concrete model  realization.
In this paper we provide a concrete model realization of these general arguments
reconsidered $S_4$ and its subgroups. We have  found   that $S_4$ is able to reproduce TBM following two different symmetry breaking patterns. We have  built two different models that  realize TBM through the two patterns dictated by the group analysis considerations and finally we discuss the possible superpotential that can break $S_4$ in the correct way.


\section{The discrete symmetry group $S_4$ as the origin of TBM}
\subsection{The group $S_4$}
The discrete group $S_4$ is given  by the permutations of four objects and it is composed by 24 elements. 
It can be defined by  two generators  $S$ and $T$ that satisfy 
\begin{equation}\label{rel}
 S^4= T^3= 1,\quad ST^2S=T \,.
 \end{equation}
The 24 elements of $S_4$ belong to five classes
\begin{eqnarray}
\label{classes}
\mathcal{C}_1&:& I \,;\nn\\
\mathcal{C}_2&:& S^2,  T S^2 T^2, S^2 T S^2 T^2 \,; \nn\\
\mathcal{C}_3&:& T,T^2, S^2 T, S^2 T^2,  S T S T^2, S T S, S^2 T S^2,S^3 T S\,;\nn\\
\mathcal{C}_4&:& ST^2,  T^2 S, T S T, T S T S^2, S T S^2, S^2 T S\,;   \nn\\
\mathcal{C}_5&:& S, T S T^2, S T, T S, S^3, S^3 T^2\,.
\end{eqnarray}
The elements of $\mathcal{C}_{2,4} $ define   two different  sets of  $Z_2$ subgroups of $S_4$,   that ones of the  class $\mathcal{C}_{4}$  a set of  $Z_3$ abelian discrete symmetries and those belonging to $\mathcal{C}_{5}$ a set of  $Z_4$ abelian discrete symmetries.
The $S_4$  irreducible representations are two singlets,  $1_1,1_2$, one doublet, $2$,  and two triplets, $3_1$ and $3_2$. We adopt the following basis 
\begin{equation}\label{base2}
S= \left(\begin{array}{cc}- 1&0\\ 0&1  \end{array} \right) \quad
T=-\frac{1}{2}\left(\begin{array}{cc} 1&\sqrt{3}\\ -\sqrt{3}& 1\end{array} \right)~,
\end{equation}
for the doublet representation and 
\bea\label{base3}
S_{+,-}= \pm \left(\begin{array}{ccc} -1&0&0\\ 0&0&-1  \\ 0&1  &0\end{array} \right) &\quad& 
T=\left(\begin{array}{ccc} 0&0&1\\ 1&0&0  \\ 0&1  &0\end{array} \right)\,,
\eea
for the triplet representations. Clearly the generators $(S_+,T)$ and  $(S_-,T)$ define the two triplet representations $3_1,3_2$ respectively.  
All the product rules can be  straightforwardly derived.  We remind the reader to the  product rules reported in \cite{Hagedorn:2006ug}.

\subsection{$S_4$ symmetry  breaking patterns}
\label{symbr}

We have seen in the introduction  that given a discrete non abelian group $G$ a predictive lepton mixing matrix may be obtained if $G$ is broken to one of its subgroups,  
with the  subgroup preserved in the charged lepton   sector  different from the subgroup preserved in the  neutrino sector. 

 We disregard therefore  the case when $S_4$ is completely broken in one of the two sectors.  At the same time,   if the left handed leptons  transform non-trivially under $S_4$, the  case of $S_4$  unbroken  in one sector is ruled out since it  leads  to a diagonal  mass matrix  with at least two degenerate states. 
Therefore  if $S_4$ is broken to one of its subgroups $G'$ in the charged lepton sector,  in the neutrino sector it has to be broken to another subgroup $G''\neq G'$. The couple $(G',G'')$  identifies a  possible symmetry breaking pattern. In this notation the lepton mixing matrix is given by
 \begin{equation}
 \label{lepgen}
 U_{lep}=U^\dag_l U_\nu= U^\dag_{G'} U_{G''}\,,
 \end{equation}
being $U_{G'},U_{G''}$ the transformations that decompose the representations of $S_4$ into the representations of $G',G''$ respectively.

    $S_4$ contains a non abelian subgroup    $S_3$,   the permutation group of three objects composed by six elements. The elements of $S_4$  that belong to $S_3$ correspond to $C_1$, $T$ and $T^2$ of  $C_3$ and $T ST S^2$,$ST S^2$, $S^2 T S$ of $C_4$.  Furthermore $S_4$ contains  the abelian subgroups $Z_2$, $Z_3$, $Z_4$ corresponding to the elements of the classes $\mathcal{C}_{2,4},\mathcal{C}_3$ and $\mathcal{C}_5$ respectively. The only representation that can break $S_4$ to $S_3$ is the triplet $3_1$.  The reason is that when a triplet  $\phi_1\sim 3_1$  develops vev as $(1,1,1)$  the six  elements that define  $S_3$ belonging to $S_4$---$ I, T, T^2, T ST S^2,ST S^2,S^2 T S$ built with the basis reported in \eq{base3}--are preserved. On the contrary, when a triplet $\phi_2\sim 3_2$  develops vev as (1,1,1), only the three elements that define $Z_3$ are preserved---$I,T,T^2$---while $T ST S^2,ST S^2,S^2 T S$ built according \eq{base3}  are broken.
 %
   
   The representations of $S_3$ are two singlet, $1_1$ and $1_2$,  and a doublet, $2$.  In general if   $S_4$ is broken  to $S_3$  the representations of $S_4$ would transform under $S_3$ according to  
\beq
 3_1\to 1_{1} + 2,\quad
 3_2\to 1_{2} + 2,\quad
 2 \to 2 ,\quad
 1_1\to 1_{1},\quad
 1_2 \to 1_{2}.
\eeq
Therefore if  $S_4$ is broken to $S_3$,   a triplet of $S_4$,  $F\sim (F_1,F_2,F_3)\sim 3_1$,  will decompose under $S_3$ as 
$F (3_1) \to \psi_0 (1_+) + \psi (2_-)$ with
\begin{equation}
 \psi_0= \frac{1}{\sqrt{3}} (F_1+F_2 +F_3),\quad
 \psi = \left(  \begin{array}{c} (F_2-F_3)/\sqrt{2} \\  (- 2 F_1  +F_2  + F_3) /\sqrt{6} \end{array} \right)\,.
\end{equation}
The new eigenstates $S_3$ $(\psi_0,\psi)$ are defined by
 \bea
 \label{newbas}
 \left(\begin{array}{c} \psi_0  \\ \psi_1 \\ \psi_2  \end{array} \right) =U_{S_3}  \left(\begin{array}{c} F_1\\ F_2  \\  F_3   \end{array} \right) & \quad \mbox{with}& \quad U_{S_3}= P\cdot  U_{TBM}^T \quad\mbox{with}\quad
 P =\left(\begin{array} {ccc} 0&1&0\\ 1&0&0 \\ 0&0&1\end{array} \right)\,.
%
 \eea

We now   assume   that $F\sim L$ being $L$  the left handed lepton doublets 
and for the moment  we leave undetermined the transformation properties under $S_4$ of the electroweak  $SU(2)$ singlets. 

The first case we  consider is the symmetry breaking pattern $(S_3, G'')$, that means that we break $S_4$ into $S_3$ in the charged lepton sector while  we still not know which is its corresponding $S_4$ subgroup in  the neutrino sector.  Assuming that the LR  charged lepton mass matrix    $M_l$  is obtained once $S_4$ is broken to $S_3$,  we can write    $M_l M_l^\dag$  in the new  basis defined by \eq{newbas}
\begin{equation}
M_l M_l^\dag \to    P \, U^T_{TBM} \,    M_l M_l^{ \dag}  \,U_{TBM} \,P=  \tilde{M}_l \tilde{M}^\dag_l\,.
\end{equation}
Since the residual symmetry is $S_3$, $\tilde{M}_l \tilde{M}^\dag_l$ has to be invariant under this symmetry. Once we impose this condition we discover that 
$  \tilde{M}_l \tilde{M}^\dag_l=M^l_{diag}  M^{l \dag}_{diag}\,,$
with $2$  degenerate masses. Neglecting  for the  moment this phenomenological inconsistency, we 
have seen that the breaking $S_4\to S_3$ in the charged lepton sector has lead to
$ U_l=  U_{TBM} P\,. $  If the neutrino mass matrix were  diagonal
$ U_{lep} =U_l^\dag U_\nu$  would lead  to the wrong conclusion  $U_{lep} =U^T_{TBM}$.  To cure this problem we have two options. On one  hand,   we could  require that the neutrino mass matrix were diagonalized by $ U_{TBM} U_{TBM} $  in order to reproduce the TBM through
$U_{lep} =  U^T_{TBM}  \, U_{TBM}\, U_{TBM}= U_{TBM}\,. $
However there is no $G''$ subgroup of $S_4$ that yields $U_{G''}= U_{TBM}\, U_{TBM}$ and therefore exact TBM cannot be obtained according to \eq{lepgen}.   On the other hand we could require to break the surviving $S_3$ in the charged lepton sector  into $Z_2$ in such a way to produce a $U_l\neq U_{TBM} P$. Even in this case there is no  corresponding $G''$ in the neutrino sector that allows to obtain exact TBM.   As consequence the  symmetry breaking pattern with  $S_4$ broken into $S_3$ in the charged lepton sector is ruled out.

\vspace{0.1cm}
 We now analyze what happens considering the breaking pattern $(Z_3,G'')$. As in the previous case the subgroup $G''$, corresponding to the neutrino sector, is  undetermined. We expect that if we  break $S_4$ into $Z_3$ in the charged lepton sector---we have already said that in $S_4$  the breaking into $Z_3$  is  realized when a triplet  $3_2$ develops a vev in the direction (1,1,1)---  the  charged lepton mixing  matrix will send the $S_4$ triplet $(L_1,L_2,L_3)$ in the $Z_3$ eigenstates, $1,1',1''$. Indeed the mixing matrix responsible of this rotation is the  $U_\omega$ defined in \eq{uom0}.
Given  $U_\omega$ the correct TBM can be reproduced if the  $U_{G''}$ of \eq{lepgen} is given by
\bea
\label{unu}
U_\nu&=& \left(\begin{array}{ccc} 0&1 &0\\ \frac{1}{\sqrt{2}}&0 & \frac{i}{\sqrt{2}} \\ \frac{1}{\sqrt{2}} &0& -\frac{i}{\sqrt{2}}  \end{array} \right)\,,
\eea
or in other words if  the neutrino mass matrix  $m^\nu$ is diagonalized by $U_\nu$  and it has the following  form
\bea
\label{mnu1}
 m^\nu &=& \left(\begin{array}{ccc} a &0&0 \\ 0 &c &b \\ 0&b&c \end{array}  \right)\,.
 \eea 
The matrix form of \eq{mnu1}  is   recovered by requiring the invariance of    $m^\nu$ under the $G''=Z_2$ subgroup of $S_4$  associated to the element  $T\,S\,T$ of the class $\mathcal{C}_4$. This breaking pattern  is the usual one used in models based on $A_4$. However we stress that in the context of $S_4$ we have obtained  TBM only according to group theory considerations.

\vspace{0.1cm}
If we consider now the case $(Z_2,G'')$ we discover that $S_4$  behaves exactly as $A_4$ and   exact TBM cannot be recovered. For a detailed analysis we remand the reader to the Appendix of  \cite{Bazzocchi:2008sp}. 

In the case $(Z_4,G'')$ we discover that  the charged lepton mass matrix  $M_l M_l^\dag$  is  diagonal with two states that are  degenerate. Since $Z_4$ is abelian this degeneration can be broken only by completely breaking $Z_4$. 
In this case $U_{G'}$ of \eq{lepgen} completely arbitrary and exact TBM cannot be obtained.

\vspace{0.1cm}
 So far we have considered all the possible cases in which  the subgroup fixed in the charged lepton sector gives rise to a non diagonal structure to  the charged lepton mass matrix $M_l$. The last case involving $Z_4$ gives rise to  a diagonal $M_l M_l^\dag$ but with two  degenerate states. We could ask if there is any way to realize a diagonal $M_l$ with three different mass eigenvalues. Indeed this is  easily  realized breaking   $S_4$  to $Z_2\times Z_2$  corresponding  to the elements  $S^2$  and  $T^2 S^2 T$ of the class $\mathcal{C}_2$.  If the charged lepton mass matrix  is diagonal all the mixing structure  arise by the neutrino sector. Therefore the last  symmetry breaking pattern  we are going to consider is $( Z_2\times Z_2, S_3)$. 
 
 In this last case we break $S_4$ into  $S_3$ in the neutrino sector.
 Following the same analysis that brought to \eq{newbas} we have 
  \beq 
  U_{TBM}^T m^\nu U_{TBM} =m^\nu_{S_3}\,,
  \eeq
 that means  $U_{lep} =U_{TBM}$ being the charged lepton mass matrix diagonal. 
At this point we have to face off a  further problem: when $S_4$ is broken to $S_3$  the  triplet $L$ splits in a singlet plus a doublet. If $S_3 $ is unbroken the two states in the doublet are degenerate in  contrast with experimental data. Therefore we should identify a way of breaking  $S_3$ without affecting the mixing rotation of the neutrino mass matrix. To keep us as general as possible, consider $m^\nu_{S_3}$ obtained once $S_4\to S_3$. If $S_3$ is unbroken we have
$
 m^\nu_{S_3}= Diag( m_1,m_0,m_0)
$\,.

 Suppose now  that the singlet and the doublet with respect to $S_3$ behave as two independent sectors in such a way that $S_3$ is preserved in the singlet sector while is broken in the doublet one\footnote{From the point of view of model realization this assumption is not different by assuming that  $S_4$ is broken to different subgroups in the charged lepton sector and  in the neutrino one. Indeed we will see in sec.~\ref{mod2} how singlet and doublet sectors  can be easily  separated. }. By imposing these conditions  we  discover that $m^\nu_{S_3broken}$ has the following expression
   \bea
 m^\nu_{S_3broken}&=& \left( \begin{array}{ccc} m_1&0& 0\\ 0&b_1&b_2 \\ 0& b_2& b_3 \end{array} \right)\,. \eea
 Finally let us impose that $S_3$ is not completely broken in the doublet sector  but it is broken  to its subgroup $Z_2$  identified by the $S_3$  generator $S$. This generator  coincides with the $S$ generator of the doublet representation of  $S_4$ given in \eq{base2}.
In this case it is possible to show that $ m^\nu_{S_3broken}= Diag( m_1,m_2,m_3)$ and 
the lepton  mixing matrix is still given by  $U_{TBM} $.
%

 We have seen that on the basis of theoretical considerations based on the subgroups of $S_4$, the flavor symmetry $S_4$ has two symmetry  breaking patterns giving exact TBM in the lepton sector. In the next section we will present a model realization for each breaking pattern. 
In the last section we  build the corresponding supepotential responsible for the correct $S_4$  symmetry breaking patterns.
 
 \section{Model realization} 
  \label{models}
  
 \subsection{Model I : $S_4 \to Z_3$ \& $S_4\to Z_2$}
\label{mod1}
 
 The first model we consider reproduces TBM through the breaking of $S_4$ into $Z_3$ and $Z_2$ in the charged lepton  and neutrino sector respectively.  We assume our model to be supersymmetric. Matter and scalar supermultiplets are reported in tab. \ref{tab1}.   The scalar supermultiplets charged under $S_4$, that in the following we will identify as flavons, are electroweak  $SU(2)\times U(1)$ singlets. Therefore the Yukawa superpotential $\mathcal{W}_Y$   of \eq{wy} includes  effective operators of order $4$. $\Lambda$ is the cutoff of the model and an extra $Z_5$ symmetry has been introduced to separate the charged lepton  sector from the neutrino one.  
 In tab. \ref{tab1} we have omitted the supermultiplets $\hat{H^u}$ and $\hat{\bar{\Phi}}$, doublet and triplet of $SU(2)$ respectively,  necessary to give mass to the up-quarks and to cancel anomalies in a  realistic model.

  \begin{table}[h!]
\begin{tabular}{|c|cc|cccccc|}
\hline
&$\hat{L}$&$\hat{E^c}$&$\hat{H^d}$&$\hat{\Phi}$&$\hat{\sigma}$& $\hat{\phi_1}$&$\hat{\phi_2}$& $\hat{\Delta}$\\ 
\hline
$SU(2)$&$2$&$1$&$2$&3&1&1&1&1\\
$S_4$&$3_1$&$3_1$&1&1&1&$3_1$&$3_2$&$3_1$\\
$Z_5$&$1$&$\omega_5^4$&1&1&$\omega_5$&$\omega_5$&$\omega_5$&1\\
\hline
\end{tabular}\caption{Matter and scalar content of   model I. The lepton mixing matrix is TB. }\label{tab1}
\end{table}
 
 The  full leading order  $S_4\times Z_5$ Yukawa superpotential $\mathcal{W}_Y$  is given by
\bea
\label{wy}
\mathcal{W}_Y&=&\frac{1}{\Lambda} y_0 (\hat{L}\hat{E^c})_{1}\hat{ \sigma} \hat{H^d}  +\frac{1}{\Lambda} y_s (\hat{L}\hat{E^c})_{3_{1}}\hat{\phi_1}\,\hat{H^d} + \frac{1}{\Lambda}\, y_a (\hat{L}\hat{E^c})_{3_{2}}\hat{\phi_2} \,\hat{H^d}+  y^\nu_1  (\hat{L} \hat{L})_{1} \,\hat{\Phi} +  \frac{1}{\Lambda}\, y^\nu_2 (\hat{L}\hat{ L})_{3_{1}} \hat{\Delta}\, \hat{\Phi}\,.
\eea
When the   $S_4$ triplet and doublet  flavons align as
\bea
\vev{\phi_1}\sim \vev{\phi_2}\sim (1,1,1) & \vev{\Delta}\sim (1,0,0)\,,
\eea
 the charged lepton and neutrino mass matrices present the usual forms
 \bea
 M_l=\left(\begin{array}{ccc} h_0 &h_1 & h_2\\ h_2 &h_0 &h_1\\ h_1&h_2&h_0    \end{array}\right) && m^\nu = \left(\begin{array}{ccc} a &0&0 \\ 0 &a &b \\ 0&b&a \end{array}  \right)
 \eea
 that satisfy
 \bea 
 U_\omega \, M_l U_\omega^\dag = M_l^{diag},\quad
 U_\nu^T \, m_\nu \,U_\nu = m_\nu^{diag}  \,,
  \eea
  with  $U_\omega$ and $U_\nu$ given in \eq{uom0} and \eq{unu} respectively.
TBM  is obtained as usual by $ U_{TB}= U_\omega U_\nu\,.$
 The mass eigenvalues for the charged lepton are given by
\begin{equation}
m_e=h_0+h_1+h_2\,,\quad m_\mu = h_0+ h_1 \omega^2 +h_2\omega\,,\quad  m_\tau = h_0+ h_1 \omega+h_2\omega^2 \,,
\end{equation}
 and for the neutrino  by $(a+b,a, b-a)$. By assuming that the flavon vevs are of order $\sim \lambda^2 \Lambda$ with $\lambda$ the Cabibbo angle,  the deviations from TBM induced by the next to leading order corrections to the Yukawa superpotential  slightly modify lepton mixing  keeping it still in agreement with neutrino data. 
Notice that the vev alignments 
\begin{equation}
\vev{\phi_1}\sim \vev{\phi_2}\sim (1,1,1) 
\end{equation} 
preserves the $Z_3$ subgroup of $S_4$ associated to the element $T$, while the vev alignments 
\begin{equation}
\vev{\varphi} \sim (0,1)\quad  \vev{\Delta}\sim (1,0,0), 
\end{equation}
preserves the $Z_2$ associated to the element $TST$ that in the doublet  and triplet  representation reads  respectively as
\begin{equation}
T\,S\,T\,=\left( \begin{array}{cc}-1&0\\ 0&1 \end{array} \right)\,, \quad  T\,S\,T\,=\left( \begin{array}{ccc}1&0&0\\ 0&0&-1\\ 0&-1&0 \end{array} \right)\,
 \end{equation} 
 
 \subsection{ Model II : $S_4\to S_3$}
\label{mod2}

The second model we describe realizes TBM through the sequential breaking of $S_4$ into $S_3$ and then into $ Z_2$ in the neutrino sector and the breaking of $S_4$  into two different  $Z_2\times Z_2$ in the charged lepton sector. The step  through $S_3$ is crucial : if we broke  $S_4$ directly into $Z_2 $ in the neutrino sector we would find a generic neutrino mass matrix $\mu-\tau$ invariant  not diagonalized  by TB. On the contrary,  in the model that we present the step through $S_3$ leads to a neutrino mass matrix   $m^\nu$ which is $\mu-\tau$ invariant and satisfy the  relation 
$ m^\nu_{11}=m^\nu_{22}+m^\nu_{23}-m^\nu_{13}\,$
that ensures TB  diagonalization. We will see that the key  ingredient in building the correct $m^\nu$ is the introduction of the right handed neutrinos transforming as a doublet of $S_4$.   As in the case of the model presented in  sec.~\ref{mod1}  we assume our model be supersymmetric and the flavon supermultiplets electroweak singlets. Matter and scalar supermultiplets are reported in tab.~\ref{tab2}.  As done in sec.~\ref{mod1} we have omitted the supermultiplet $\hat{\bar{\Phi}}$, triplet of $SU(2)$, necessary to cancel anomalies. Two extra discrete abelian symmetries, $Z_3$ and $Z_5$,  have been introduced in order to avoid interferences between the sectors.

\begin{table}[h!]
\begin{tabular}{|c|ccc|ccccccc|}
\hline
&$\hat{L}$&$\hat{l^c}$&$\hat{N^c}$&$\hat{H^u}$&$\hat{H^d}$&$\hat{\Phi}$&$\hat{\Delta}$&$\hat{\sigma}$& $\hat{\phi}$&$\hat{\varphi}$\\
\hline
$SU(2)$&$2$&$1$&1&$2$&2&3&1&1&1&1\\
$S_4$&$3_1$&$3_1$&2&1&1&$1$&$3_1$&1&2&2\\
$Z_3$&$\omega^2$&1&1&1&$\omega^2$&$\omega$&$\omega$&1&1&1\\
$Z_5$&$1$&$\omega^3_5$&1&1&1&1&1& $\omega^2_5$&$\omega^2_5$&1\\
\hline
\end{tabular}\caption{Matter and scalar content of   model II. The lepton mixing matrix is TB. }\label{tab2}
\end{table}
The full leading order $S_4\times Z_3\times Z_5$  invariant Yukawa  superpotential is given by
\bea
\mathcal{W}_Y&=&\frac{1}{\Lambda} y_s (\hat{L}\hat{l^c})_{1} \hat{\sigma}  \hat{H^d}+\frac{1}{\Lambda}  y_d (\hat{L}\hat{l^c})_{2}\hat{\phi}\,\hat{H^d} + y_1 (\hat{L}\hat{L})_1 \hat{\Phi} + \frac{1}{\Lambda}y_2 (\hat{L}\hat{ \Delta})_2 \hat{N^c}  \hat{H^u}  + M_d \hat{ N^c}\hat{ N^c} + \tilde{y}_N\,\hat{\varphi} \hat{N^c}\hat{ N^c}\,,
\eea
where as usual $\Lambda$ is the cutoff of the model and all  the Yukawa terms are of order 4 with the exception of the ones involving right handed neutrinos.
 We assume that the  flavons  $\Delta$  and $\varphi$, triplet and doublet under $S_4$ respectively,  align  as 
\bea
\vev{\Delta}\sim (1,1,1) &\quad& \vev{\varphi}\sim(0,1)\,.
\eea
The vev $\vev{\Delta}$ preserves $S_3$ as has been already discussed in  sec.~\ref{symbr}. The vev  $\vev{\varphi}$ preserves  the $S$ generators of $S_3$ 
that coincides with the $S$ generator of $S_4$ of the doublet representation---\eq{base2}.

The doublet $\phi$ does not align and  develops vev as $\vev{\phi}\sim (v_1,v_2)$---this means that $S_4$  is broken  to $Z_2\times Z_2$ corresponding to the elements $S^2$ and $ T\, S^2 T^2$ of $\mathcal{C}_2$ that in the $3_1$  triplet representation read as 
\beq S^2 = \left( \begin{array}{ccc} 1&0&0\\ 0& -1&0\\ 0&0&-1\end{array}\right)\,,\quad  T\, S^2 T^2= \left( \begin{array}{ccc} -1&0&0\\ 0& 1&0\\ 0&0&-1\end{array}\right)\,. \eeq

For the charged lepton sector we have 
\bea
M_l &=& \frac{1}{\Lambda} v^d  \left( \begin{array}{ccc} y'_s v_\sigma - 2 y''_d v^\phi_2 &0&0 \\ 0&y'_s v_\sigma+ y'_d v^\phi_1+ y''_d v^\phi_2 &0 \\ 0&0& y'_s v_\sigma-y'_d v^\phi_1+y''_d v^\phi_2 \end{array}  \right)\,
\eea
with 
$v_\sigma= \vev{\sigma} \quad v^\phi_{1,2} =\vev{\phi_{1,2}}\quad v^d=\vev{H^d_0}\,,$
and the product factors absorbed in $y_s'$ and $y_d',y_d''$.
The neutrino mass matrix gets  contributions both  from type I and type II see-saw
\beq m^\nu= m_{LL}-m_D\cdot \frac{1}{M_N} \cdot m_D^T\,, \eeq
where  $m_{LL}=  y_1 v_\Phi \cdot \mathcal{I}$  with $v_\Phi=\vev{\Phi}$ and 
\bea
m_D= y_2 \frac{v^\Delta}{\Lambda} v^u \left( \begin{array}{cc} 0&-2\sqrt{6}   \\
 1/\sqrt{2}   &1/\sqrt{6}    \\
  -1/\sqrt{2}  &1/\sqrt{6}    \end{array}  \right)\,,
  &\quad&
  M_N= \left( \begin{array}{cc}  M_d+V_\varphi&0\\ 0&M_d -V_\varphi \end{array} \right)\,,
\eea
with  $v^u=\vev{H^u_0}$,  $v_{1,2,3}^\Delta=v^\Delta$ and $ V_\varphi= \tilde{y}_N \vev{\varphi_2}/\sqrt{2}$.  After the usual see-saw mechanism the  majorana neutrino mass matrix is given by   
\bea
m^\nu &=&  \left( \begin{array}{ccc} a+ \frac{2}{3} b&  -\frac{1}{3} b& -\frac{1}{3} b\\
   -\frac{1}{3} b& a+  \frac{ 1 }{6} b+\frac{1}{2} c& \frac{ 1 }{6} b -\frac{1}{2} c\\ -\frac{1}{6} b& \frac{ 1 }{6} b  -\frac{1}{2} c& a  +\frac{ 1}{6} b  +\frac{1}{2} c\end{array}  \right)\,,
\eea
with
\beq a   =y_1 v_\Phi\,, \quad b=-y_2^2 \left (\frac{v^\Delta}{\Lambda}\right)^2 \frac{( v^u )^2}{M_d-V_\varphi} \,,\quad c=-y_2^2\left (\frac{v^\Delta}{\Lambda}\right)^2 \frac{ (v^u )^2}{M_d+V_\varphi}\,.\eeq
The neutrino mass matrix $m^\nu$ is diagonalized by TBM  
and its eigenvalues are $(a+b,a, a+c)$ that can accommodate experimental neutrino mass splitting data  being expressed in terms of three independent combinations of the parameters of the model. As in the model discussed in sec.~\ref{mod1} by assuming the flavon vevs of order $\sim \lambda^2 \Lambda$ next to leading order corrections to the Yukawa superpotential produce  small deviations from TBM that are  still compatible with neutrino data.


\section{Realizing the correct vacuum configurations in $S_4$}

In the context of flavor model based on non abelian discrete symmetry the lepton TBM is obtained thanks to specific alignments of the flavons. The so-called alignment problem in $A_4$ and $T'$ has been extensively discussed in \cite{Ma:2006vq,Altarelli:2005yx,Altarelli:2005yp}.  Different strategies have been used:  the introduction of soft breaking term of the flavor symmetry \cite{Ma:2006vq},  the use of  a continuous $U(1)_R$ symmetry \cite{Altarelli:2005yx} preserved by the scalar potential and the promotion of the model to a fifth dimension \cite{Altarelli:2005yp}. In the context of $S_4$ in \cite{Koide:2007sr} the flavon superpotential was softly broken to guarantee the desired vacuum configuration.

In   $S_4$ as well as in $A_4$ and $T'$   it is impossible to build a flavon  superpotential  that guarantees  the  alignments needed.  In the next sections we will show that  the extra discrete abelian symmetries  introduced in  sec.~\ref{models}  to separate the two  lepton sectors are sufficient to give  the correct vacuum configurations. 

\subsection{Model I : minimization of the  potential}
\begin{table}[h!]
\begin{tabular}{|c|cccc|ccc|}
\hline
&$\hat{\sigma}$& $\hat{\phi_1}$&$\hat{\phi_2}$& $\hat{\Delta}$&$\hat{\varphi}$& $\hat{\xi}$&$\hat{\eta}$\\ 
\hline
$SU(2)$&1&1&1&1&1&1&1\\
$S_4$&1&$3_1$&$3_2$&$3_1$&2&2&2\\
$Z_5$&$\omega_5$&$\omega_5$&$\omega_5$&1&1&$\omega_5^3$ &$\omega_5^2$\\
\hline
\end{tabular}\caption{Scalar content of   model I including the flavons that contribute to the mass matrix structures and the ones the drive the correct vacuum alignments, the driving fields.}\label{scalar1}
\end{table}
The flavon potential is obtained by the following part of the full  $S_4\times Z_5$ superpotential
\bea
\label{potmod1}
\mathcal{W}_Y &=& M_{\xi\eta}\hat{\xi} \hat{\eta}+ \lambda_{\xi \eta }\hat{\xi} \hat{\eta}\hat{\varphi}+  \lambda_{\sigma\eta }\hat{\sigma} \hat{\eta} \hat{\eta}+ \lambda_{\xi\phi1}   \hat{\xi} \hat{\phi_1} \hat{\phi_1} + \lambda_{\xi\phi2}   \hat{\xi} \hat{\phi_2} \hat{\phi_2} + \lambda_{\xi\phi12}   \hat{\xi} \hat{\phi_1} \hat{\phi_2} \nn\\
&+&M_{\Delta }\, \hat{\Delta}\hat{\Delta}  + M_{\varphi}\, \hat{\varphi}\hat{\varphi} + \lambda_{\varphi \Delta}\, \hat{\Delta}\hat{\Delta} \hat{\varphi} +  \lambda_{\varphi}\, \hat{\varphi}\hat{\varphi} \hat{\varphi}+\lambda_\Delta\, \hat{\Delta}\hat{\Delta}\hat{\Delta} \,.
\eea
We  assume that the  flavor symmetry is broken in the SUSY limit and therefore the vacuum configuration  is obtained solving the system
$
{\partial \mathcal{W}_Y}/{\partial f_i}=0\,,
$
where $f_i$ are the $f$ components of the supermultiplets entering in \eq{potmod1} and $i$ runs on all the  supermultiplets.
 By assuming the general vacuum configuration
%
\beq
\vev{\Delta} = (v^\Delta_1,v^\Delta_2,v^\Delta_3),\, 
\vev{\varphi} =(v^\varphi_1,v^\varphi_2),\, 
\vev{\phi_1}= (v^\phi_1,v^\phi_2,v^\phi_3),\,  
\vev{\phi_2}= (u^\phi_1,u^\phi_2,u^\phi_3),\,
\vev{\xi}= (u^\xi_1, u^\xi_2),\, 
\vev{\eta}= (z^\eta,z^\eta)~\vev{\sigma}=v_\sigma,
\eeq
the  set of equations  is given by
 \bea
a)~ 
\partial W/\partial f^\Delta_1&=&
\frac{2}{\sqrt{3}}M_\Delta  v^\Delta_1 -\frac{2}{\sqrt{3}}\lambda_{\Delta \varphi} v^\Delta_1 v^\varphi_2+ 2  \lambda_\Delta v^\Delta_2 v^\Delta_3  =0\nn\\
b)~  
{\partial W}/{\partial f^\Delta_2}&=&  
\frac{2}{\sqrt{3}}M_\Delta  v^\Delta_2 + \frac{1}{\sqrt{3}}\lambda_{\Delta \varphi} v^\Delta_2 (v^\varphi_2+ \sqrt{3} v^\varphi_1) + 2  \lambda_\Delta v^\Delta_1 v^\Delta_3  =0\nn\\
c)~   
{\partial W}/{\partial f^\Delta_3}&=& 
\frac{2}{\sqrt{3}}M_\Delta  v^\Delta_3 + \frac{1}{\sqrt{3}}\lambda_{\Delta \varphi} v^\Delta_3 (v^\varphi_2- \sqrt{3} v^\varphi_1) + 2  \lambda_\Delta v^\Delta_1 v^\Delta_2  =0\nn\\
d)~    
{\partial W}/{\partial f^\varphi_1}&=
&\sqrt{2} M_\varphi  v^\varphi_1 +\frac{\lambda_{\xi\eta}}{2} ( u^\xi_2 z^\eta_1 + u^\xi_1 z^\eta_2)+\frac{\lambda_\Delta}{2}[ (v^\Delta_2)^2 -(v^\Delta_3)^2]=0\nn\\
e)~   
{\partial W}{\partial f^\varphi_2}
&=&\sqrt{2} M_\varphi  v^\varphi_2 +\frac{\lambda_{\xi\eta}}{2} ( u^\xi_1 z^\eta_1 - u^\xi_2 z^\eta_2)+\frac{\lambda_\Delta}{2\sqrt{3}}[-2 (v^\Delta_1)^2+ (v^\Delta_2)^2 +(v^\Delta_3)^2]=0\nn\\
f)~    
{\partial W}/{\partial f^{\eta}_1}&=&
\frac{M_{\xi \eta}}{\sqrt{2}} u^\xi_1 +\frac{\lambda_{\xi \eta} }{2}(v^\varphi_1 u^\xi_2+ v^\varphi_2 u^\xi_1) + \sqrt{2} \lambda_{\sigma \eta} v_\sigma z^\eta_1 =0\nn\\
g)~  
{\partial W}/{\partial f^{\eta}_2}&=&
\frac{M_{\xi \eta}}{\sqrt{2}} u^\xi_2 +\frac{\lambda_{\xi \eta} }{2}(v^\varphi_1 u^\xi_1 - v^\varphi_2 u^\xi_2) + \sqrt{2} \lambda_{\sigma \eta} v_\sigma z^\eta_2  =0\nn\\
 h)~     
{\partial W}/{\partial f^{\sigma}}
&=&\frac{\lambda_{\sigma\eta}}{\sqrt{2}} [(z_1^\eta )^2+ (z_2^\eta )^2]  =0\nn\\
i)~     
{\partial W}/{\partial f^{\xi}_1}
&=&\frac{1}{\sqrt{2}}M_{\xi \eta} z^\eta_1 + \frac{1}{2} \lambda_{\xi \eta} ( z^\eta_1 v^\varphi_2+ z^\eta_2 v^\varphi_1  )   +   \frac{1}{2}  \lambda_{\xi \phi1} [(v^\phi_2)^2 -(v^\phi_3)^2] +  \frac{1}{2}  \lambda_{\xi \phi2} [(u^\phi_2)^2 -(u^\phi_3)^2] \nn\\
&+& \frac{1}{2\sqrt{3}}  \lambda_{\xi \phi12} (2 v^\phi_1 u^\phi_1- v^\phi_2 u^\phi_2- v^\phi_3 u^\phi_3)=0   \nn\\
j)~       
{\partial W}/{\partial f^{\xi}_2}&=& 
\frac{1}{\sqrt{2}} M_{\xi \eta} z^\eta_2 +  \frac{1}{2} \lambda_{\xi \eta} ( z^\eta_1 v^\varphi_1 - z^\eta_2 v^\varphi_2  ) +      \frac{1}{2\sqrt{3}}  \lambda_{\xi \phi1} [-2 (v^\phi_1)^2+(v^\phi_2)^2 +(v^\phi_3)^2] \nn\\ &+&  \frac{1}{2\sqrt{3}}  \lambda_{\xi \phi2} [-2 (u^\phi_1)^2+(u^\phi_2)^2 +(u^\phi_3)^2]+        \frac{1}{2}  \lambda_{\xi \phi12} (v^\phi_2 u^\phi_2- v^\phi_3 u^\phi_3)=0\nn\\
k)~   
{\partial W}/{\partial f^{\phi_1}_1}&=& 
\frac{1}{\sqrt{3}} (\lambda_{\xi\phi12} u^\phi_1 u^\xi_1-2 \lambda_{\xi \phi1} u^\xi_2 v^\phi_1)  =0\nn\\ 
l)~     
{\partial W}/{\partial f^{\phi_1}_2}&=& 
u^\xi_1 ( \lambda_{\xi \phi1} v^\phi_2 -\frac{1}{2\sqrt{3}}  \lambda_{\xi\phi12} u^\phi_2) +  u^\xi_2   (\frac{ \lambda_{\xi \phi1}}{\sqrt{3}} v^\phi_2 +\frac{1}{2} \lambda_{\xi\phi12}  u^\phi_2)      =0\nn\\ 
m)~   
{\partial W}/{\partial f^{\phi_1}_3}&=&
u^\xi_1 (- \lambda_{\xi \phi1} v^\phi_3 -\frac{1}{2\sqrt{3}}  \lambda_{\xi\phi12} u^\phi_3) +  u^\xi_2   (\frac{ \lambda_{\xi \phi1}}{\sqrt{3}} v^\phi_2 -\frac{1}{2} \lambda_{\xi\phi12} u^\phi_2)   =0\nn\\
n)~    {\partial W}/{\partial f^{\phi_2}_1}&=&  \frac{1}{\sqrt{3}} (\lambda_{\xi\phi12} v^\phi_1 u^\xi_1-2 \lambda_{\xi \phi2} u^\xi_2 u^\phi_1) =0\nn\\ 
o)~      {\partial W}/{\partial f^{\phi_2}_2}&=&  u^\xi_1 ( \lambda_{\xi \phi2} u^\phi_2 -\frac{1}{2\sqrt{3}}  \lambda_{\xi\phi12} v^\phi_2) +  u^\xi_2   (\frac{ \lambda_{\xi \phi1}}{\sqrt{3}} u^\phi_2 +\frac{1}{2} \lambda_{\xi\phi12}  v^\phi_2)  =0\nn\\ 
   p)~     {\partial W}/{\partial f^{\phi_2}_3}&=&  u^\xi_1 (- \lambda_{\xi \phi 2} u^\phi_3 -\frac{1}{2\sqrt{3}}  \lambda_{\xi\phi12} v^\phi_3) +  u^\xi_2   (\frac{ \lambda_{\xi \phi2}}{\sqrt{3}} u^\phi_2 -\frac{1}{2} \lambda_{\xi\phi12} v^\phi_2) =0 \label{eqpot1}
 \eea
 Eq.~$h)$ of  \eq{eqpot1}  implies $z^\eta_{1,2}=0$.  As first consequence we have that a possible solution of    eqs.$f)-g)$ and eqs. $k)- p)$   is given by
\begin{equation}
(u_1^\xi,u_2^\xi)=(0,0)\quad \mbox{and} \quad  v_\sigma\neq 0 \,.
\end{equation}
 By substituting 
 $ (z^\eta_1,z^\eta_2)=(0,0) $, $ (u_1^\xi,u_2^\xi)=(0,0)$ and $ v_\sigma\neq 0$
 in the equations not yet solved 
it is easy to check that  a possible solution for eqs.~$a)-e)$ is given by the vacuum configuration
 \bea
 (v_1^\varphi,v_2^\varphi) =(0,v^\varphi) &\mbox{with}& v^\varphi=\frac{M_\Delta}{\lambda_\Delta}\nn\\
 (v_1^\Delta,v_2^\Delta,v_3^\Delta) =(v^\Delta,0,0) &\mbox{with}& v^\Delta=  6^{1/4} \frac{ \sqrt{M_\varphi M_\Delta}}{\lambda_\Delta}\,.
 \eea
Finally eqs.~$i)-j)$ are solved by the  vacuum configuration
 \bea
 (v_1^\phi,v_2^\phi, v_3^\phi)=v^\phi (1,1,1) &\mbox{and} & (u_1^\phi,u_2^\phi, u_3^\phi)=u^\phi (1,1,1) \,.
 \eea
The solution found is not unique but can be stabilized once we add apposite  SUSY soft breaking terms. In sec.~\ref{mod1} we have assumed that the flavon vevs is of order $\lambda^2 \Lambda$. Therefore the next to leading order corrections to the Yukawa superpotential induced by the driving fields are sufficiently suppressed.  

\subsection{Model II : minimization of the potential}

\begin{table}[h!]
\begin{tabular}{|c|ccccc|cc|}
\hline
&$\hat{\Delta}$&$\hat{\sigma}$& $\hat{\phi}$&$\hat{\varphi}$&$\hat{\bar{\sigma}}$&$\hat{\xi}$&$\hat{\eta}$\\
\hline
$SU(2)$&1&1&1&1&1&1&1\\
$S_4$&$3_1$&$1$&2&2&1&1&1\\
$Z_3$&$\omega$&1&1&1&1&$\omega$&$\omega^2$\\
$Z_5$&1& $\omega^2_5$&$\omega^2_5$&1&$\omega_5$&1&1\\
\hline
\end{tabular}\caption{Scalar content of   model I including both  flavon and the driving field supermultiplets. }\label{scalars2}
\end{table}
The flavon potential is obtained by the following part of the full superpotential
\bea
\mathcal{W} &=& \lambda_{\Delta \xi}\hat{\xi} \hat{\Delta} \hat{\Delta}+\lambda_\Delta  \hat{\Delta} \hat{\Delta} \hat{\Delta} +  M_\xi  \hat{\xi} \hat{\eta} + \lambda_\xi  \hat{\xi}  \hat{\xi}  \hat{\xi} + \lambda_{\eta} \hat{\eta}  \hat{\eta}  \hat{\eta}  \nn\\
&+& M_\varphi \hat{\varphi} \hat{\varphi}    +\lambda_\varphi   \hat{\varphi} \hat{\varphi}   \hat{\varphi} +
\lambda_\phi \hat{\bar{\sigma}} \hat{\phi} \hat{\phi} + \lambda_\sigma \hat{\bar{\sigma}} \hat{\sigma} \hat{\sigma}\,.
\eea
By assuming the general vacuum configuration
\begin{equation}
\vev{\Delta}= (v^\Delta_1,v^\Delta_2,v^\Delta_3),\, 
\vev{\varphi} =(v^\varphi_1,v^\varphi_2) ,\, 
\vev{\phi} =(v^\phi_1,v^\phi_2) ,\, 
\vev{\xi}=v_{\xi}  ,\, 
\vev{\eta}=v_{\eta}~ \vev{\sigma}=v_\sigma,\,  
\vev{\bar{\sigma}}=v_{\bar{\sigma}},
\end{equation}
%
%
the minimization of the scalar potential obtained in the SUSY limit gives the following set of equations
\bea
\label{minpote2}
{\partial  \mathcal{W}_Y}/{\partial f^\Delta_1}&=&\sqrt{2} \lambda_{\Delta \xi} v_\xi v^\Delta_1+\sqrt{3} 3 \lambda_\Delta   v^\Delta_2  v^\Delta_3=0\nn\\
{\partial  \mathcal{W}_Y}/{\partial f^\Delta_2}&=&  \sqrt{2}  \lambda_{\Delta \xi} v_\xi v^\Delta_2+ \sqrt{3} \lambda_\Delta   v^\Delta_1  v^\Delta_3=0\nn\\
{\partial  \mathcal{W}_Y}/{\partial f^\Delta_3}&=& \sqrt{2}  \lambda_{\Delta \xi} v_\xi v^\Delta_3+ \sqrt{3}  \lambda_\Delta   v^\Delta_1  v^\Delta_2=0\nn\\
{\partial \mathcal{W}_Y}/{\partial f ^\xi}&=&\sqrt{3} \lambda_{\Delta \xi} [(v_1^\Delta)^2 + (v_2^\Delta)^2 + (v_3^\Delta)^2 ] + M_\xi v_{\eta}+ 3 \lambda_\xi v_\xi^2 =0\nn\\
{\partial  \mathcal{W}_Y}/{\partial f ^{\eta}}&=& M_\xi v_\xi+ 3 \lambda_{\eta} v_{\eta}^2 =0\nn\\
{\partial  \mathcal{W}_Y}/{\partial f ^\varphi_1}&=&  \sqrt{2} M_\varphi \,v^\varphi_1+ 3 \lambda_\varphi v^\varphi_1 v^\varphi_2=0\nn\\
{\partial  \mathcal{W}_Y}/{\partial f ^\varphi_2}&=&  \sqrt{2} M_\varphi \,v^\varphi_2+ \frac{3}{2} \lambda_\varphi [(v^{\varphi}_1)^2 -(v^{\varphi}_2)^2]=0\nn\\
{\partial  \mathcal{W}_Y}/{\partial f ^\phi_1}&=&  \sqrt{2}  \lambda_{\phi} v^\phi_1 v_{\bar{\sigma}}=0\nn\\
{\partial  \mathcal{W}_Y}/{\partial f ^\phi_2}&=&  \sqrt{2}  \lambda_{\phi} v^\phi_2 v_{\bar{\sigma}}=0\nn\\
{\partial  \mathcal{W}_Y}/{\partial f ^\sigma}&=& 2 \lambda_{\bar{\sigma}} v_\sigma \,v_{\bar{\sigma}} =0\nn\\
{\partial  \mathcal{W}_Y}/{\partial f ^{\bar{\sigma}}}&=&\frac{1}{\sqrt{2}} \lambda_\phi [(v_1^\phi)^2+ (v_2^\phi)^2 ]+\lambda_{\bar{\sigma}} \, v_\sigma^2=0\,.
\eea
Discarding for the triplet and the doublets  the trivial solutions that do not break $S_4$, the solution of the system of  \eq{minpote2} is given by the following vacuum configuration
\bea
v^\Delta_1=v^\Delta_2=v^\Delta_3= v^\Delta& \mbox{with}& v^\Delta=\sqrt{2}\frac{\lambda_{\Delta \xi} \lambda_{\eta}}{\lambda_\Delta}\frac{v_{\eta}^2}{M_\xi} \nn\\
v_\xi= -3 \lambda_{\eta} \frac{v_{\eta}^2}{M_\xi} & \mbox{with}& v_{\eta}^3=- M_\xi^3 \frac{\lambda_\Delta^2}{\lambda_{\eta}^2 (2\sqrt{3} \lambda_{\Delta \xi}^3 + 27 \lambda_\xi \lambda_\Delta^2)}\nn\\
(v^\varphi_1,v^\varphi_2) \neq (0,0) & \mbox{with} & \left \{ \begin{array}{c} (0,\frac{2 \sqrt{2}}{3} \frac{M_\varphi}{\lambda_\varphi})\\ 
(\sqrt{\frac{2}{3}}\frac{M_\varphi}{\lambda_\varphi},  -\frac{\sqrt{2}}{3}\frac{M_\varphi}{\lambda_\varphi} )\\
(-\sqrt{\frac{2}{3}}\frac{M_\varphi}{\lambda_\varphi},   -\frac{\sqrt{2}}{3}\frac{M_\varphi}{\lambda_\varphi} )
 \end{array}\right. \nn\\
v_\sigma^2 = -\frac{1}{\sqrt{2}}\frac{\lambda_\phi}{\lambda_\sigma}  [ (v^\phi_1)^2 +  (v^\phi_2)^2&]\neq 0~\mbox{and}&
 v_{\bar{\sigma}}=0\,.
\eea
The three solutions corresponding to $\vev{\varphi}$ are degenerate and corresponding to the breaking of $S_3$ to its  $3$  different  $Z_2$ subgroups.  Through  appropriate choices of soft terms that break the discrete abelian symmetry $Z_3$ and $Z_5$ and not $S_4$ we can stabilize as absolute minimum the vacuum configuration $\vev{\varphi}\sim(0,1) $.

\section{Conclusion}

In this paper we have discussed the idea that $S_4$ is the minimal discrete non abelian group naturally related to TBM in the lepton sector.  We have shown that $S_4$ can yield exact TBM according to a general group theory analysis and we have presented   two explicit model realizations of how TBM can be obtained in $S_4$ once the basis of its generators are fixed. In addition we have  provided a detailed study of the corresponding scalar potentials.
The two models require two triplets  with different vev alignments. 
For each  model we have built a potential that in the SUSY limit contains the minimum required.   The problem of the triplet and doublet alignments 
is solved in a more economical way than  in models based on $A_4$ \cite{Ma:2001dn,Babu:2002dz,Hirsch:2003dr,Ma:2004zv,Altarelli:2005yp,Chen:2005jm,Zee:2005ut,Altarelli:2005yx,
Adhikary:2006wi,Valle:2006vb,Adhikary:2006jx,Ma:2006vq,Altarelli:2006kg,Hirsch:2007kh,Altarelli:2007cd,Bazzocchi:2007na}.
%
%
To separate the charged lepton  sector from  the neutrino one we have introduced extra abelian symmetries.
The construction of the potentials have not required additional symmetries than such extra abelian symmetries, but just 
the addition of ``driving''  fields that do not enter in the Yukawa part.
We have  studied  neither the quark sector  nor 
the possibility to embed such a model in a GUT theory. We leave these subjects  for a future publication. 
It is worth to  mention that in $S_4$ there is more freedom to generate the mixing in the quark sector than in $A_4$. Indeed the doublet irreducible 
representation  could play an important role as happens  in $T'$ \cite{Feruglio:2007uu}.

\section*{Acknowledgments}

We thank L. Merlo for very useful comments and discussions. Work  supported by 
MEC grant 
FPA2008-00319/FPA, by EC RTN network MRTN-CT-2004-503369
and by Generalitat Valenciana ACOMP06/154.

 \def\baselinestretch{1}


\end{document}